\documentclass[twocolumn,usenatbib]{mn2e}
\usepackage{graphicx}
\usepackage{natbib}
\usepackage{bm}
\usepackage{amsfonts}
\usepackage{latexsym}
\usepackage[latin1]{inputenc}
\usepackage{amsmath}
\usepackage{epsfig}
\usepackage{amsbsy}
\usepackage{mnfixes}

\newcommand{\mtb}[1]{\mathbf{#1}}

\newcommand{\mvt}[1]{\emph{\textbf{#1}}}

\newcommand{\n}{\mathrm n}
\newcommand{\p}{\mathrm p}

\newcommand{\x}{\mathrm x}
\newcommand{\y}{\mathrm y}

\newcommand{\rmd}{\mathrm d}

\newcommand{\veps}{\varepsilon}

\def \nn  {\nonumber}

\def \veps{\varepsilon}

\newcommand{\beq}{\begin{equation}}
\newcommand{\eeq}{\end{equation}}

\newcommand{\brac}[1]{\left({#1}\right)}

\newcommand{\pd}[2]{\frac{\partial{#1}}{\partial{#2}}}
\newcommand{\td}[2]{\frac{\rmd{#1}}{\rmd{#2}}}
\newcommand{\curl}{\nabla\times}

\def\jnl@style{\it}
\def\aaref@jnl#1{{\jnl@style#1}}

\def\aaref@jnl#1{{\jnl@style#1}}

\def\aj{\aaref@jnl{AJ}}                   
\def\apj{\aaref@jnl{ApJ}}                 
\def\apjl{\aaref@jnl{ApJ}}                
\def\apjs{\aaref@jnl{ApJS}}               
\def\apss{\aaref@jnl{Ap\&SS}}             
\def\aap{\aaref@jnl{A\&A}}                
\def\aapr{\aaref@jnl{A\&A~Rev.}}          
\def\aaps{\aaref@jnl{A\&AS}}              
\def\mnras{\aaref@jnl{MNRAS}}             
\def\prd{\aaref@jnl{Phys.~Rev.~D}}        
\def\prc{\aaref@jnl{Phys.~Rev.~C}}        
\def\prl{\aaref@jnl{Phys.~Rev.~Lett.}}    
\def\qjras{\aaref@jnl{QJRAS}}             
\def\skytel{\aaref@jnl{S\&T}}             
\def\ssr{\aaref@jnl{Space~Sci.~Rev.}}     
\def\zap{\aaref@jnl{ZAp}}                 
\def\nat{\aaref@jnl{Nature}}              
\def\aplett{\aaref@jnl{Astrophys.~Lett.}} 
\def\apspr{\aaref@jnl{Astrophys.~Space~Phys.~Res.}} 
\def\physrep{\aaref@jnl{Phys.~Rep.}}      
\def\physscr{\aaref@jnl{Phys.~Scr}}       

\title[Stratification, superfluidity and magnetar QPOs]{Stratification, superfluidity and magnetar QPOs}

\author[A. Passamonti $\&$ S. K. Lander]
{A. Passamonti\thanks{E-mail: andrea.passamonti@uni-tuebingen.de} , S. K. Lander \thanks{E-mail: samuel.lander@uni-tuebingen.de}
\\ \\  
Theoretical Astrophysics, IAAT, Eberhard Karls University of T\"{u}bingen, T\"{u}bingen 72076, Germany}

\begin{document}

\date{\today}

\pagerange{\pageref{firstpage}--\pageref{lastpage}} \pubyear{}

\maketitle

\label{firstpage}


\begin{abstract}
The violent giant flares of magnetars excite QPOs which persist for hundreds
of seconds, as seen in the X-ray tail following the initial
burst. Recent studies, based on single-fluid barotropic magnetar
models, have suggested that the lower-frequency QPOs
correspond to magneto-elastic oscillations of the star. The higher 
frequencies, however --- in particular the strong 625 Hz peak
--- have proved harder to explain, except as high mode multipoles. 
In this work we study the time evolutions of non-axisymmetric oscillations of two-fluid
Newtonian magnetars with no crust. We consider models with superfluid neutrons and normal protons, 
and poloidal and toroidal  background field configurations. 
We show that multi-fluid physics (composition-gradient stratification, entrainment)  
 tends to increase Alfv\'en mode frequencies significantly from
their values in a single-fluid barotropic model. The higher-frequency
magnetar QPOs may then be naturally interpreted as Alfv\'en oscillations of the multi-fluid stellar
core. The lower-frequency QPOs are less easily explained within our
purely fluid core model, but we discuss the possibility that these are crustal modes. 
\end{abstract}


\section{Introduction} \label{sec:intro}

The quasi-periodic oscillations (QPOs) of magnetars provide a
tantalising possibility of probing the interior physics of neutron
stars.  Detected in the aftermath of giant flares, these QPOs are
thought to relate directly to oscillation modes of the underlying
star, perturbed by the hugely energetic flare. The theoretical
challenge is to provide sufficiently good models of magnetar
oscillations that the observed QPOs may be convincingly linked to
particular modes; this could provide potential constraints on the
stellar equation of state and magnetic field configuration.

Magnetars are thought to be slowly-rotating and magnetically-powered
neutron stars (NSs), with dipole surface values reaching $10^{15}$ G
--- an exceptional value, even for neutron stars.   Three of these objects have been seen to produce enormously energetic `giant
flares': SGR 0526-66~\citep{1983A&A...126..400B}, 
SGR 1900+14 and SGR 1806-20~\citep{2005ApJ...628L..53I, 2005ApJ...632L.111S, 2006ApJ...637L.117W}. 
In the latter two cases, QPOs were detected in the flare's
decaying X-ray tail, between $100$ and $400$ seconds after the initial
burst. A number of frequencies were identified within each tail, of
varying duration and intensity. 

Initially these QPOs were identified with crustal shear modes, and
there was hope of using these to constrain the equation of
state~\citep{1998ApJ...498L..45D,2005ApJ...628L..53I}. Since these
optimistic early studies, the picture has become more complex, due to
the effect of the magnetic field: the crust and core are coupled,
with modes becoming magneto-elastic in
character~\citep{2006MNRAS.368L..35L, 2006MNRAS.371L..74G,2012MNRAS.421.2054G}. Attempts to model observed QPOs
as global modes were dealt a blow by~\citet{2007MNRAS.377..159L}, who
used a toy model to show that axisymmetric oscillations of a magnetic
star could form a continuum of frequencies, with discrete crustal modes quickly damped
and QPOs representing the edges of the Alfv\'en continuum. Numerical
simulations supported this conclusion in the axial
case~\citep{2008MNRAS.385L...5S}, but 
found that polar oscillations were discrete modes, with no evidence
for a continuum~\citep{2009MNRAS.395.1163S}. These two studies
considered axisymmetric oscillations on a purely poloidal background
field.

There are now indications that situations with coupling between axial and polar
sectors may destroy the continuum. Two 
studies of non-axisymmetric magnetar oscillations --- for purely
toroidal and purely poloidal background fields --- found only discrete
modes~\citep{2010MNRAS.405..318L,2011MNRAS.412.1730L}. These seemed
at odds with the results for axisymmetric studies; 
the key difference may be that non-axisymmetric Alfv\'en modes are not purely
axial or purely polar\footnote{Instead, they are `axial-led' or `polar-led',
using the terminology of~\citet{1999ApJ...521..764L}.}. For
axisymmetric modes, a purely poloidal background field has decoupled
axial and polar modes, but a mixed poloidal-toroidal field couples the
two sectors. These coupled axial-polar oscillations were also recently found to be
discrete modes~\citep{2012MNRAS.423..811C}.

Despite the great increase in sophistication of magnetar QPO
modelling, a major missing ingredient in all these studies has been the
multi-fluid, non-barotropic nature of neutron star
matter. A large fraction of mature neutron stars (including magnetars)
is composed of superfluid neutrons and superconducting
protons~\citep{1969Natur.224..673B,2012MNRAS.422.2632H}, and the
resulting oscillation spectrum will surely be different from that of
a barotropic star. Furthermore, there are 
indications that no stable magnetic equilibria exist in barotropic
stars~\citep{2009AA...499..557R,2012MNRAS.tmp.3162L} --- in which case
it is clearly undesirable to use them as  
background models when studying oscillations. The reason why many
magnetar QPO studies have not encountered these instabilities so far is that
they specialise to axisymmetric oscillations on poloidal-field
backgrounds; the strongest instabilities appear for non-axisymmetric
perturbations~\citep{1973MNRAS.163...77M,1973MNRAS.162..339W}.

In this work we try to advance the modelling of magnetar QPOs, by
studying the oscillations of a stratified multi-fluid star. We consider poloidal and toroidal magnetic
field geometries. Two main simplifications we make are that the whole star
is fluid (with no crust), and that the neutrons are superfluid but the
protons normal; the validity of these assumptions is discussed in the
following sections. 

We begin by reviewing in Section~\ref{sec:Eq} the main formalism for studying the dynamics of a magnetised two-fluid star. 
First we determine the background configurations and then we derive the linearised dynamical equations. 
In Section~\ref{sec:results} we present our results, where we explore the differences from barotropic
single-fluid magnetar models. Finally in Section~\ref{sec:discussion}, we discuss our results in the
context of magnetar QPOs and potential future improvements.

\section{Equation of Motion} \label{sec:Eq}

The basic matter constituents of a neutron star are neutrons, protons
and electrons (npe), although more exotic particle species may exist
in an inner core region. In this work we assume a npe-composition in the entire volume of the star and neglect the inner core.  

From a dynamical point of view, we may assume that the core's protons and electrons can be considered as a single co-moving fluid,  as the electromagnetic interaction 
locks them on timescales much smaller than the dynamical oscillation periods. 
The dynamics of this system can be therefore described by a two-fluid model comprising a gas of superfluid neutrons and 
a neutral mixture of protons and electrons that we call for simplicity  ``protons''.   
Neutron and proton quantities will be indicated with a subscript roman n and p,  respectively.

The equations  to study the dynamics of a superfluid and magnetised star are, in the ideal MHD approximation,  the mass and momentum conservation equations 
of each fluid constituent, the induction equation for the magnetic field, and the Poisson equation for  the gravitational potential. These equations read~\citep{2011MNRAS.410..805G}:
\begin{align}
& \partial_t \rho_{\x} + \nabla  \cdot \left( \rho_{\x} \mvt{v}_{\x} \right)
= 0 \, , \label{eq:Mcon} \\
& \left( \partial_t + \mvt{v}_{\x} \nabla  \right)
 \left( \mvt{v}^{\x} + \varepsilon_{\x}  \mvt{w}_{\y\x} \right) + \nabla \left( \Phi + \tilde{\mu}_{\x} \right)
+ \varepsilon_{\x} w_{k}^{\y\x} \nabla  v_{\x}^{k}  =  \frac{ \mvt{F}_{\x}}{\rho_{\x}} \, ,  \label{eq:Euler} \\
& \partial_{t} \mvt{B} = \nabla \times  \left(    \mvt{v}_{\p} \times  \mvt{B} \right)  , \label{eq:Beq} \\
& \nabla^2 \Phi = 4 \pi G \rho \,  , \label{eq:Poisson} 
\end{align}
where the labels x and y (with $\x \neq \y$) denote the fluid component n and p. 
The quantities $\rho_{\x}$, $\tilde \mu _{\x}$ and $\mvt{v}_{\x}$ are, respectively,  the mass density, chemical potential 
and velocity of each fluid constituent, while  the relative velocity is denoted by 
 $\mvt{w}_{\x\y} = \mvt{v}_{\x} - \mvt{v}_{\y}$.  The gravitational potential and the magnetic field are described by 
 $\Phi$ and $\mvt{B}$, respectively, and $\rho = \rho_{\n} + \rho_{\p}$ is the total mass density. 
 
The parameter $\veps_{\x}$ accounts for the entrainment between nucleons, which is a non dissipative effect 
that in neutron stars is due to the strong interaction. The main effect of entrainment is to couple the nucleon motion by 
inducing a relative dragging between neutrons and protons.  As a result  the conjugate momentum of each component 
is not aligned  with its velocity,  see equation~(\ref{eq:Euler}). 
The entrainment parameter can be also written in terms of the nucleon's effective mass $m_{\x}^{\star}$ 
by using the relation  $\veps_{\x} = 1 - m_{\x}^{\star} / m_{\x}$~\citep{2002A&A...393..949P}.

The vector field $\mvt{F}_{\x}$ in equation~(\ref{eq:Euler}) represents the force density that acts on the $\x$ fluid component. 
The magnetic interaction  and the mutual friction are the forces expected in a magnetised superfluid star with no crust.  
For normal (not superconducting) protons the interaction with the magnetic field is given by the 
Lorentz force:  
\begin{equation}
\mvt{F}_{L} = \frac{1}{4\pi}  \left( \nabla \times \mvt{B}  \right) \times   \mvt{B}  \, , \label{eq:fL}  
\end{equation}
where we have used Amp\`{e}re's law to replace the charge density current.  
The mutual friction is instead a dissipative force mediated by superfluid vortices and operates in rotating stars on both the fluid components. 
In this work we are interested in the oscillation spectrum of magnetars, which are slowly rotating objects that can be 
 well described by non-rotating stellar models.  We can therefore neglect the effects of rotation and mutual friction.  
More details on the impact of mutual friction on the oscillation spectrum in unmagnetised stars can be found in~\cite{2011MNRAS.413...47P}.  

Under these assumptions, the force density in  equation~(\ref{eq:Euler}) is given by 
 $\mvt{F}_{\p} = \mvt{F}_{L}$ and $\mvt{F}_{\n} = 0$.

\subsection{Equation of State} \label{sec:EoS}

The equation of state (EoS) can be described by an energy functional
\begin{equation}
\mathcal{E} = \mathcal{E} \left( \rho_\n, \rho_\p , w_{\n \p}^2
\right) \, , \label{eq:EoS}
\end{equation}
that ensures Galilean invariance.  The chemical potential
$\tilde{\mu}_\x$ and the entrainment parameter $\varepsilon_{\x}$ are
then defined by
\begin{eqnarray}
\tilde{\mu}_{\x} & \equiv & \left. \frac{\partial \mathcal{E}}{\partial \rho_{\x} }
\right| _{\rho_{\y}, w_{\x\y}^2}\, , \label{eq:defmu} \\
 \varepsilon_{\x} & \equiv & 2 \rho _{\x}\left.  \frac{\partial
\mathcal{E}}{\partial w^{2}_{\n\p} } \right|_{\rho_\x,\rho_\y} \, . \label{eq:vareps}
\end{eqnarray}
If the relative velocity between the two fluids is small, a typical situation in most astrophysical systems,  
equation~(\ref{eq:EoS}) can be expanded in series
\begin{equation}
\mathcal{E} = \mathcal{E}_{0} \left(\rho_\n, \rho_\p \right)
+ \alpha_0 \left( \rho_\n, \rho_\p \right) w_{\n \p}^2 + \mathcal{O}\left(w_{\n \p}^4\right) \,  ,  \label{eq:EoSbulk}
\end{equation}
and the bulk EoS $\mathcal{E}_{0}$ and the entrainment parameter
$\alpha_0$ can 
be independently specified at $\mtb{w}_{\n\p}=\mtb{0}$.  
This approximation is certainly valid in our star's  model as a non zero relative velocity appears 
at first perturbation order. 
From equation~(\ref{eq:vareps}) it follows that the entrainment parameter
$\varepsilon _{\x}$ is given by $ \rho_\x  \varepsilon_\x = 2 \alpha_0. $

We consider an analytical EoS, which is a two-fluid analogue of a polytropic 
model~\citep{2002A&A...393..949P, 2002PhRvD..66j4002A,
  2009MNRAS.396..951P, 2012MNRAS.419..732L}:
\begin{equation}
\mathcal{E}_{0} = k_{\n} \, \rho_\n^{\gamma_n} + k_{\p} \,
\rho_\p^{\gamma_\p} \, ,  \label{eq:EosPR}
\end{equation}
where the coefficients $k_\x$ are constants and $\gamma_\x$ is related to the adiabatic index $N_{\x}$ by the 
standard definition $\gamma_{\x}=1+1/N_{\x}$. Despite its simplicity, this EoS allows us to construct 
 neutron star models with composition gradients when $N_{\n} \neq N_{\p}$.  
 Stratification is a relevant property for the oscillation spectrum and may be crucial for the 
 stability of the background magnetic configuration.

\subsection{Background models: magnetic two-fluid stellar equilibria} \label{sec:back_pert}

We begin by modelling a neutron star as an axisymmetric body in
Newtonian gravity. For this system it is natural to use cylindrical polar coordinates
$(\varpi,\phi,z)$, where the $z$-axis is aligned with the symmetry axis
of the star's magnetic field.

We assume the whole interior of the magnetar is multi-fluid, with the
neutrons superfluid, but the protons normal rather
than superconducting. Although this is done for simplicity, it may be
realised in magnetars if their internal field strength reaches $\sim
10^{16}$ G; at this `upper critical field' superconductivity is
broken~\citep{2011MNRAS.410..805G}. We do not account for the presence
of the neutron star crust in this study.

Since a full description of the equilibrium equations and their solution are
presented in~\citet{2012MNRAS.419..732L}, we provide only a brief
summary of the important details here. We neglect the inertia of the
electrons and absord their chemical potential into that of the
protons, so our equations describe a two-fluid system. 
From equation~\eqref{eq:Euler}, we see that the separate Euler equations governing the two
fluids are:
\begin{align} 
& \nabla\brac{\tilde\mu_\n+\Phi}=0,  \label{n_Euler} \\ 
& \nabla\brac{\tilde\mu_\p+\Phi}=\frac{(\curl \mvt{B} )\times  \mvt{B} }{4\pi\rho_\p}. \label{p_Euler} 
\end{align}
As in the single-fluid case we have Poisson's
equation~(\ref{eq:Poisson}) for the gravitational potential, from which we see that the
behaviour of the two fluids is coupled, despite the neutrons being a superfluid.

The equilibrium magnetic field of an axisymmetric barotropic star is
governed by a single equation in terms of the magnetic streamfunction
$u$, known as the Grad-Shafranov equation. Although our stellar models
are multi-fluid and stratified, our chosen equation 
of state still allows us to derive a variant of the usual
Grad-Shafranov equation, by replacing the total density $\rho$ with
the proton-fluid density $\rho_\p$:
\beq \label{GS_eq}
\brac{\pd{^2}{\varpi^2}-\frac{1}{\varpi}\pd{ }{\varpi}+\pd{^2}{z^2}}u
 = -4\pi\rho_\p\varpi^2\td{M}{u} - f\td{f}{u} \, , 
\eeq
where $M=M(u)$ is a scalar function related to the Lorentz force and
$f=f(u)$ dictates the structure of the toroidal component in
a mixed-field configuration. 
This equation describes equilibria with purely poloidal and mixed
poloidal-toroidal fields. We defer mixed fields to future
work, and therefore set $f(u)=0$ to produce pure-poloidal field
models. Pure-toroidal field models are not governed by a separate
equation but just involve an extra term in the proton-Euler equation.

We employ an iterative numerical scheme to self-consistently solve
the equations \eqref{eq:Poisson} and \eqref{eq:EosPR}-\eqref{GS_eq} in integral form; this produces our background equilibria.

\subsection{Perturbation Equations}

The perturbation equations for studying magnetised superfluid stars can be derived by linearising equations~(\ref{eq:Mcon})-(\ref{eq:Beq}) 
and using as dynamical variables the mass density, velocity and magnetic field  perturbations. However, we prefer to define a new set of 
perturbation variables which both reflects the dynamical degrees of freedom of a two-fluid system and simplifies the implementation of the 
boundary conditions. A pulsating two-fluid star behaves dynamically as a coupled harmonic oscillator in which we may discern a co- and counter-moving 
relative motion between the two constituents. In general, these two degrees of freedom are coupled, but in some particular cases 
they can be completely decoupled --- for instance in a non-stratified stellar model with no magnetic field. 

Following \citet{2009MNRAS.396..951P}, we use equations~(\ref{eq:Mcon})-(\ref{eq:Euler}) to derive the following system of dynamical equations for non-rotating stars: 
\begin{eqnarray}
 \partial_{t}  \mvt{f} & = & - \nabla \delta P   + \frac{\nabla P }{\rho} \delta \rho  -\rho \nabla \delta \Phi   + \rho_{\p} \delta \left(  \mvt{F}_{L} /  \rho_{\p} \right) \, ,  \label{eq:dfdt} \\
   \partial_{t}  \mvt{D}   & = &  \gamma_{\veps} ^{-1}  \left( 1 - x_{\p} \right)  \rho_{\p} \left[ - \nabla \delta \beta   + \delta \left(   \mvt{F}_{L} / \rho_{\p} \right) \right]  \, , \label{eq:dDdt}  \\
 \partial_{t}  \delta \rho & = & - \nabla \cdot \mvt{f} \, ,  \label{eq:drho} \\
 \partial_{t} \delta \chi_{\p} & = & - \nabla \cdot  \mvt{D}  - \mvt{f} \cdot \nabla x_{\p} \, . \label{eq:dchip}
\end{eqnarray}
where we have defined $ \gamma_{\veps} =  1 - \veps_{\n} - \veps_{\p} $ and the following fluid perturbation variables:
\begin{eqnarray}
 \mvt{f} & = & \rho_{\n} \delta\mvt{v}_{\n} + \rho_{\p} \delta\mvt{v}_{\p}  \, , \\
 \delta \rho & = &\delta \rho_{\n} + \delta \rho_{\p} \, , \\
 \mvt{D} & = &\rho_{\p} \left( 1- x_{\p} \right) \delta \mvt{w}_{\p\n}  \, ,  \\
 \delta \chi_{\p} & = &\rho \,  \delta x_{\p}  \, . 
\end{eqnarray}
The quantity $\delta P$ is the total pressure perturbation that for a co-rotating background reads 
\begin{equation}
\nabla \delta P = \delta \left( \rho_{\n} \nabla \tilde \mu_{\n} + \rho_{\p} \nabla \tilde \mu_{\p}  \right) \, ,
\end{equation}
while $\delta \beta = \delta \tilde \mu_{\p} - \delta \tilde \mu_{\n}$ describes the deviation from beta equilibrium induced by oscillations and 
$\delta \Phi$ is the gravitational potential perturbation.

To simplify the boundary conditions  we choose a ``flux'' variable also for the magnetic field perturbation, i.e. we define  $\mvt{b} = \rho_{\p} \delta \mvt{B}$. The 
 induction equation then reads
\begin{align}
\partial_t \mvt{b} & = x_{\p}  \nabla \times \left( \mvt{f} \times \mvt{B} \right) + \nabla \times \left( \mvt{D} \times \mvt{B} \right)  - x_{\p}   \frac{\nabla \rho}{\rho} \times \left( \mvt{f} \times \mvt{B} \right)  \nn \\
&  - \frac{\nabla \rho_{\p} }{ \rho_{\p} } \times \left(  \mvt{D} \times \mvt{B} \right)  \, , \label{eq:dB}
\end{align}
while the perturbation of the Lorentz force in equations~(\ref{eq:dfdt}) and (\ref{eq:dDdt})  assumes the following form:
\begin{equation}
\delta \mvt{F}_{L}  = \left(  \nabla \times \mvt{B} \right) \times \frac{  \mvt{b} }{\rho_{\p} } +  \left(  \nabla \times \mvt{b} \right) \times \frac{  \mvt{B} }{\rho_{\p} } 
- \frac{ 1 } {\rho_{\p}^2 } \left(  \nabla \rho_{\p}  \times \mvt{b} \right) \times \mvt{B} \, .
\end{equation}

It is worth noticing that the Alfv\'en velocity in a two-fluid model is given by~\citep*{2009MNRAS.396..894A} 
\begin{equation}
v_{A} ^{2} = \frac{B^2}{4 \pi \rho_{\p} }  \,  ,   \label{eq:vA}
\end{equation}
which depends on the proton density instead of the total mass density as in the single-fluid case.

The perturbation of the gravitational potential  is obtained by solving the linearised Poisson equation 
\begin{equation}
\nabla \delta \Phi = 4 \pi G \delta \rho \, . \label{eq:dPois}
\end{equation}

Due to the symmetry of our background configurations, any perturbation variables may be Fourier expanded  
with respect to the coordinate $\phi$. 
More precisely, in orthonormal spherical coordinates the mass density  as well as any other perturbation function can be written in the following form
~\citep{2002MNRAS.334..933J}
\begin{equation}
\delta \rho  = \sum_{m=0}^{\infty}
               \left[ \delta \rho_{ m}^{+} \left( t,r,\theta\right)
               \cos m \phi + \delta \rho_{m}^{-} \left(
               t,r,\theta\right) \sin m \phi \right] \, , 
               \label{eq:drhoexp}
\end{equation}
where $m$ is the azimuthal  harmonic  index. With this transformation the perturbation equations 
decouple with respect to $m$ and the problem becomes two-dimensional in 
$\left( r, \theta \right)$.

\subsubsection{Boundary conditions}

 For non-rotating stars  with purely poloidal or toroidal magnetic fields, the two-dimensional numerical domain  
extends over the region $0\leq r \leq R$ and $0\leq \theta \leq \pi/2$, where $R$ is the stellar radius. 
To study the evolution of non-axisymmetric oscillations ($m\neq 0$) we must therefore 
specify the boundary conditions at  the star's origin,  rotation axis,
equator and surface. In particular, for $m\geq 2$
 the regularity of the linearised equations leads  
to a zero condition for all the scalar and vector perturbation fields at the origin ($r=0$) and rotation axis ($\theta=0$), e.g. $\delta \rho = 0$.   
At the equator ($\theta=\pi/2$), the reflection symmetry splits the perturbation variables  in two sets of opposite parity~\citep{2002MNRAS.334..933J}.  
If we represent all the scalar and vector perturbation fields by, respectively, the mass density and mass flux perturbations,  
the polar-led perturbations satisfy the following conditions at the equator:
\begin{align}
\partial_{\theta} \delta \rho = \partial_{\theta}  f^{r} = f^{\theta} = \partial_{\theta}  f^{\phi} = 0 \, . 
\end{align}
The conditions for the axial-led class are
\begin{align}
 \delta \rho = f^{r} = \partial_{\theta} f^{\theta} =   f^{\phi} = 0 \, . 
\end{align}
The boundary conditions for the magnetic field perturbations depend on
the background field configuration \citep{2010MNRAS.405..318L}. For a poloidal field we have
\begin{align} \label{polar-poloidal}
b^r = \partial_{\theta} b^\theta = b^\phi = 0
\end{align}
for polar-led perturbations and
\begin{align} \label{axial-poloidal}
\partial_{\theta} b^r = b^\theta = \partial_{\theta} b^\phi = 0
\end{align}
for the axial-led class. For a toroidal field, by contrast, the
\emph{polar-led} perturbations obey condition \eqref{axial-poloidal} and the \emph{axial-led}
perturbations obey \eqref{polar-poloidal}.

In our purely fluid models, i.e. with no crust, the two fluid components extend 
over the entire stellar volume. The crust is an important constituent 
of a neutron star which we intend to implement in a future work. 
In the current study the position of the star's surface ($r=R$) is given by the vanishing of the proton and neutron mass density, i.e.  
$\rho_{\p} = \rho_{\n} = 0$.
For the perturbation variables we require that the Lagrangian perturbation of the chemical potentials vanish at the surface, i.e. 
\begin{equation}
\Delta \tilde \mu_{\x} = \delta \tilde \mu_{\x} + \xi_{\x}  \cdot \nabla \tilde \mu_{\x} = 0 \,  , 
\end{equation} 
where $\xi_{\x}$ is the Lagrangian displacement associated with each fluid constituent~\citep{2004MNRAS.355..918A}. 
These conditions may be expressed in terms of pressure and chemical potential perturbations as 
$\Delta P = \Delta \beta = 0$. For polytropic models, the surface condition may be further simplified to $\delta P = 0$~\citep{2009MNRAS.396..951P}.   
From the definition of the stellar surface ($\rho_{\x} = 0$) it follows that all the ``flux'' variables are zero at $r=R$, i.e. $\mvt{f} = \mvt{D} = \mvt{b} = \mvt{0}$.

\subsubsection{Numerical Code}

We study the time evolution of equations~(\ref{eq:dfdt})-(\ref{eq:dchip}) and $(\ref{eq:dB})$ with a numerical code based on a 
MacCormack algorithm. The code is an extension of those already developed for a superfluid star by~\cite{2009MNRAS.396..951P}   and 
a magnetised star by~\cite{2010MNRAS.405..318L}.  In these two references the reader can find all the technical details. 
In order to speed up the numerical simulation we neglect the perturbation of the gravitational potential and adopt the so-called 
Cowling approximation. In this way we do not have to solve the elliptic equation~(\ref{eq:dPois}) which is time consuming. 
This approximation has anyway a tiny effect on the Alfv\'en mode frequencies and it is suitable for the aims of this work.  

We consider for the axial- and polar-led perturbations the same initial conditions used by~\cite{2010MNRAS.405..318L} and\cite{2011MNRAS.412.1730L}.

\begin{figure}
\begin{center}
 \includegraphics[height=75mm]{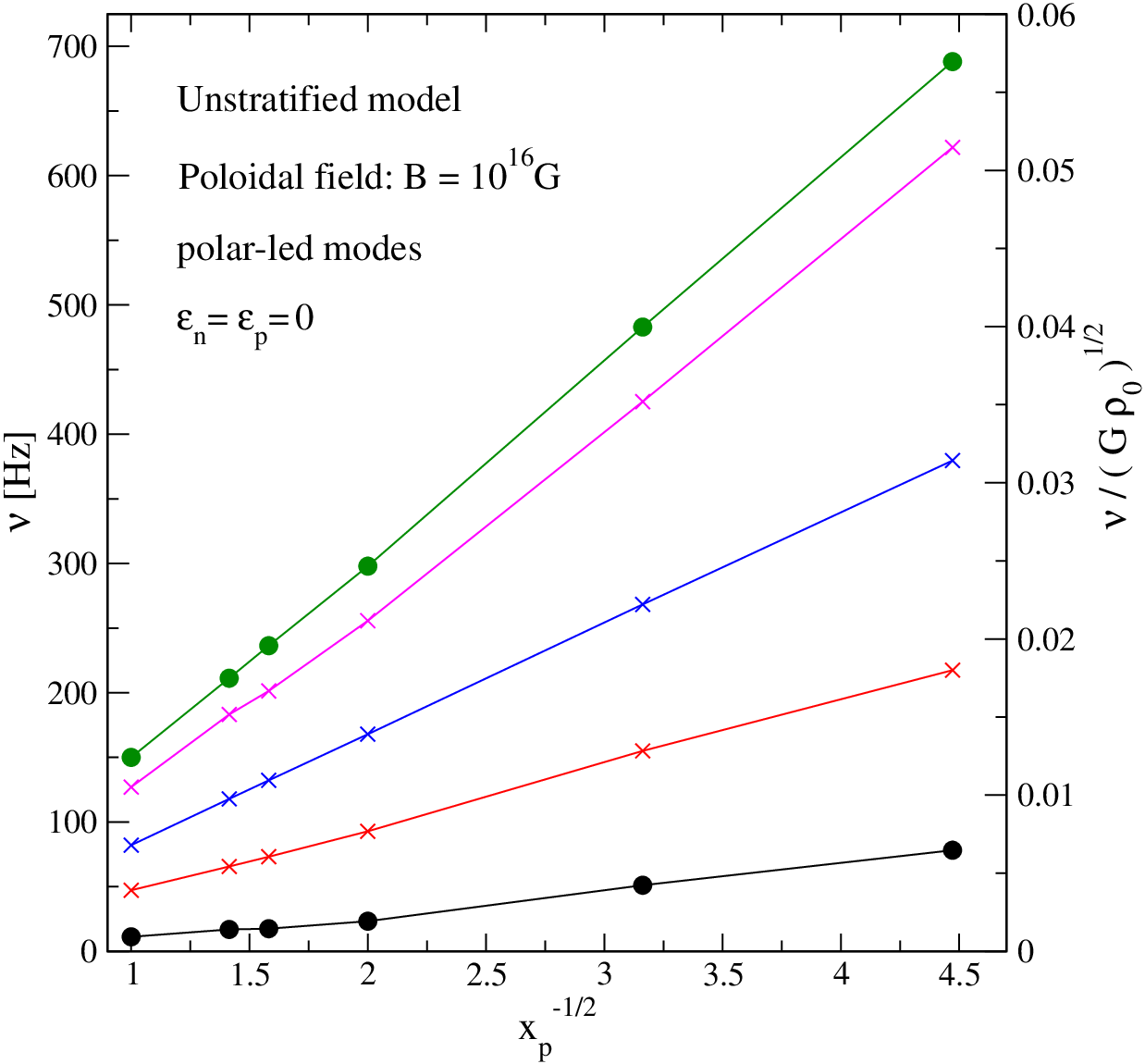}
\caption{Polar-led $m=2$ Alfv\'en modes for an unstratified star with 
a purely poloidal magnetic field with average magnitude $B=10^{16}$~G. 
This figure shows the dependence of mode frequency on the proton fraction for a
stellar model with zero entrainment. On the left vertical  axis, the mode frequency $\nu = \sigma / (2 \pi)$ is shown 
in physical units for a star with $M=1.4M_{\odot}$ and $R=10 \, \textrm{km}$. The right vertical axis instead shows the 
dimensionless mode frequencies  $\nu / ( G \rho_0 )^{1/2}$, where $\rho_0$ is the central mass density of the background model. 
 In the limit $x_\p\to 1$ we recover 
the single-fluid results. The modes already determined (not determined) in the single-fluid limit by~\citet{2011MNRAS.412.1730L}  
are denoted with a cross (filled-circle). 
\label{fig1} }
\end{center}
\end{figure}

\section{Results} \label{sec:results}

We are interested in studying the effects of two-fluid physics on magnetar dynamics. 
A realistic magnetar model has complex physics which includes superfluid/superconducting 
components, crust, realistic EoS, a strong magnetic field and magnetosphere. 
In this work we make a first step toward the introduction of superfluid physics in magnetised stars,  
and study its impact on the Alfv\'en mode spectrum. Here, we focus on purely fluid models, while  the presence of a crust 
and the study of magneto-elastic oscillations will be addressed in future work. 

We consider both stratified and unstratified stellar models which can be obtained with an appropriate 
choice of the polytropic indices in the EoS~(\ref{eq:EosPR}). More precisely, models with 
constant (non-constant) proton fraction can be determined by setting $N_{\p} = N_{\n}$ ($N_{\p} \neq N_{\n}$). 
For the unstratified case, we choose $N_\n=N_\p=1$ and study the effects of entrainment and proton fraction on 
the oscillation mode frequencies.  
For the stratified case, we construct models with an increasing composition-gradient stratification 
and study its impact on the spectrum.

This work studies non-axisymmetric oscillations of stars
with purely poloidal and purely toroidal fields. We
are able to investigate both axial- and polar-led modes of a
toroidal-field star, but the axial-led perturbations of a
poloidal field are subject to the rapidly-growing Tayler instability
\citep{1973MNRAS.162..339W,1973MNRAS.163...77M,2011MNRAS.412.1730L}, 
which dominates the evolutions. In this case we consider  only polar-led modes. 
We specialise to the case of oscillations of azimuthal index $m=2$ for brevity. 
Whilst it is possible to study modes of higher $m$ using our code, we
anticipate that they will be similarly affected by two-fluid physics;
in addition, higher-$m$ modes are probably more
susceptible to damping in a real magnetar and hence less likely to
survive to produce the observed long-lived QPOs.

\subsection{Oscillations of unstratified magnetars}

We begin by studying unstratified models, which are described by the EoS~(\ref{eq:EosPR}) with 
$N_{\p}=N_{\n}=1$.

From the plane-wave 
analysis of~\citet{2009MNRAS.396..894A},  we expect the Alfv\'en mode frequencies to scale as 
\begin{equation}
\sigma = \sqrt{ \frac{ \veps_{\star} }{ x_{\p}} } \, \sigma_0 \label{eq:pw} \, , 
\end{equation}
where $\sigma_0$ is the mode frequency of a single-fluid star, and  $\veps_{\star}$ is given by  
 \begin{equation}
\veps_{\star} = \frac{ 1 - \veps_{\n} } { 1 - \veps_{\n} - \veps_{\p} } \, .
\end{equation}
The presence of the proton fraction $x_{\p}$ in equation~(\ref{eq:pw}) is due to the different 
definition of the Alfv\'en velocity in a two-fluid system, as shown in equation~(\ref{eq:vA}).
Equation~(\ref{eq:pw}) will guide the analysis of our numerical results, and we also expect  that 
in the appropriate `single-fluid' limits (in terms of $x_\p$, $\veps_\star$) we should recover
the mode frequencies reported in \citet{2010MNRAS.405..318L} for a
background toroidal field and \citet{2011MNRAS.412.1730L} for a
background poloidal field. 

\begin{figure}
\begin{center} 
\includegraphics[height=75mm]{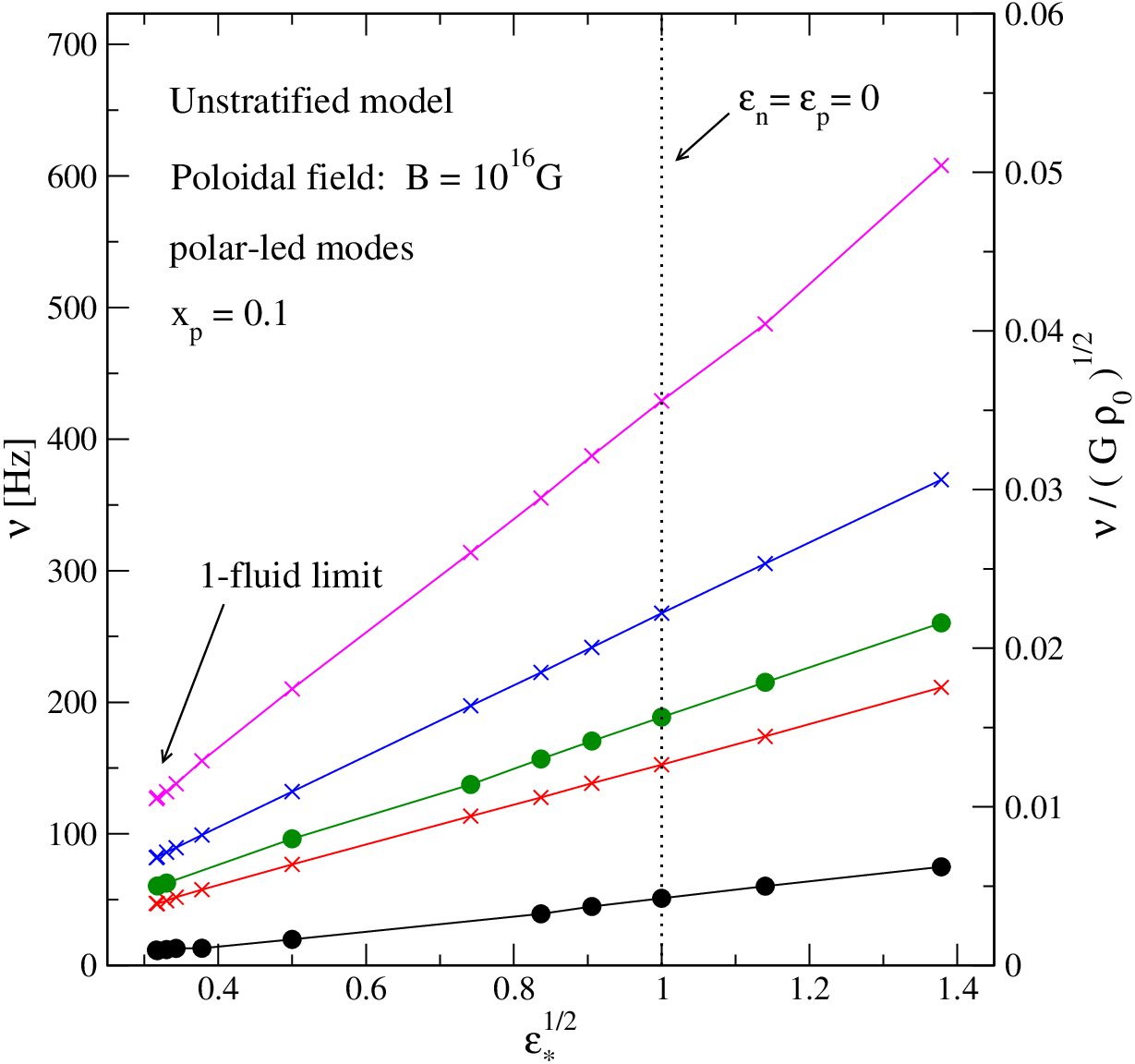}
\caption{Polar-led Alfv\'en $m=2$ modes for an unstratified star with 
a purely poloidal magnetic field with average magnitude
$B=10^{16}$~G. This figure displays the mode dependence on the entrainment $\veps_{\star}$ for a star with a constant proton fraction $x_\p = 0.1$. 
A realistic parameter space for the core's entrainment without strong pinning would be $1\leq \veps_{\star}^{1/2}  \leq  1.7$ (see text).  
In the limit of large  effective masses $m_{\x}^{\star} \gg m_\x$, i.e.  $\veps_{\star} \simeq x_{\p}$,  the mode frequencies tend as expected to the single-fluid results 
determined by~\citet{2011MNRAS.412.1730L}.  The notation used in this figure is 
the same as in figure~\ref{fig1}. \label{fig2}}
\end{center}
\end{figure}

For all the results reported here for unstratified stars, we take a
fiducial average field strength of $10^{16}$ G, which corresponds to a
polar-cap value of $5\times 10^{15}$ G for a poloidal field. This is a
little larger than observed magnetar fields, but the correspondingly shorter Alfv\'en
timescale allows for faster numerical evolutions. We have, however,
checked that for all our models the mode frequencies exhibit the
expected linear scaling with magnetic field strength.

In figure \ref{fig1} we plot the dependence of polar-led Alfv\'en mode frequencies
on the proton fraction $x_\p$, for a star with a
purely poloidal field and zero entrainment.
In this and following figures, the mode frequencies $\nu=\sigma / (2 \pi)$  are shown both in dimensionless units, $\nu / (G \rho_0)^{1/2}$ (where $\rho_0$ is the central mass density),  
and in physical units, for which  we have considered  a star with mass $M = 1.4 M_{\odot}$ and radius $R=10$~km. 
It is evident from figure~\ref{fig1} that the expected relationship from the plane-wave analysis, i.e.  $\sigma\propto
x_\p^{-1/2}$,  is borne out by our full numerical
analysis.  
 This is not surprising as in a non-stratified model the relevant parameters are constant. 
Most notably, we are able to determine the spectrum from a small proton fraction, $x_{\p} = 0.05$,  up to 
the single-fluid case $x_{\p}=1$, where we recover the results for a 
single-fluid magnetised star.

\begin{figure}
\begin{center}
\includegraphics[height=74mm]{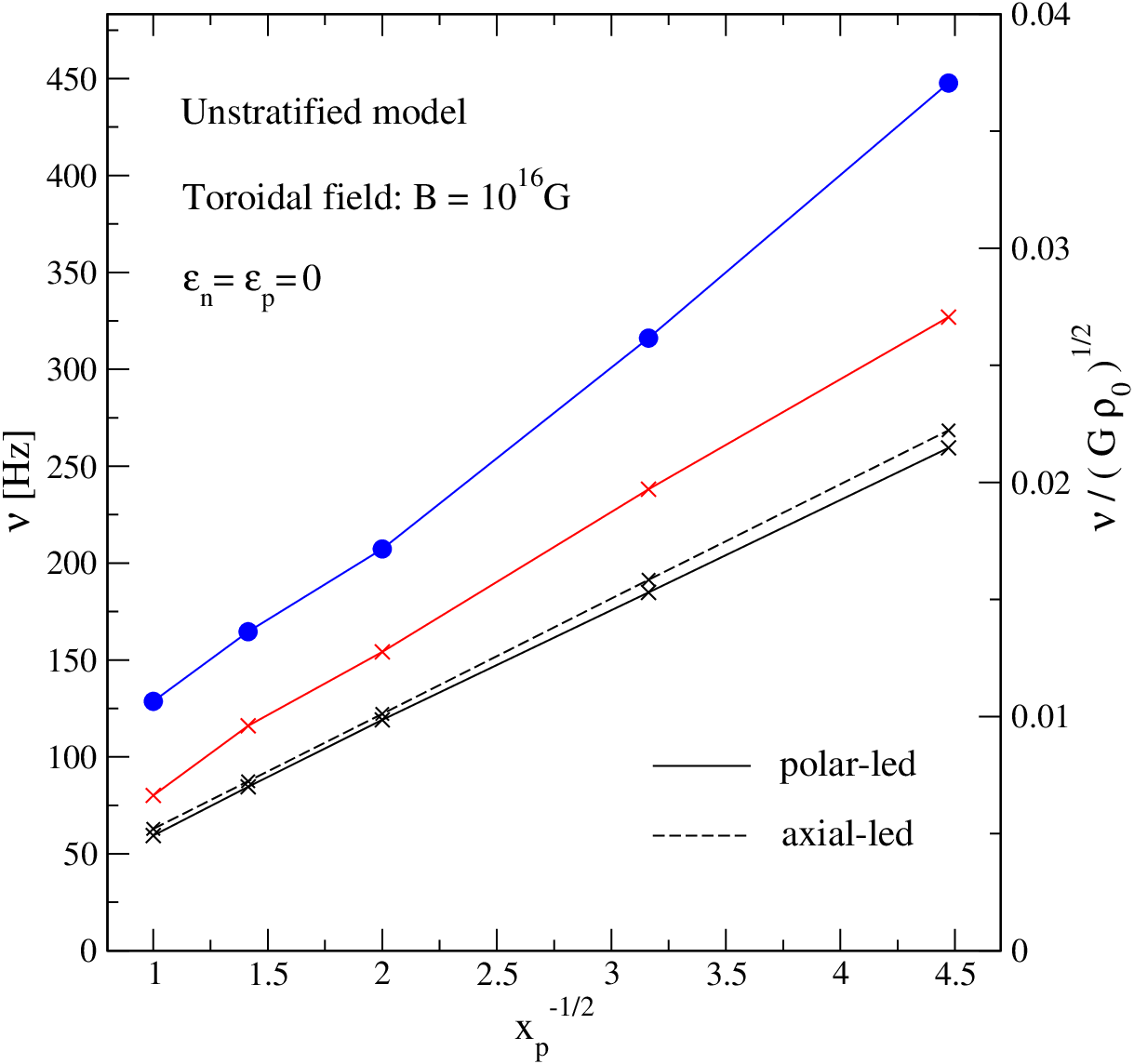}
\caption{ Axial- and polar-led $m=2$ Alfv\'en modes for an unstratified star with
  purely toroidal magnetic field with $B=10^{16}$~G and zero entrainment. This figure shows the
  modes' scaling with proton fraction. The mode frequencies determined by~\citet{2010MNRAS.405..318L} 
 are recovered in the single-fluid limit. See figure~\ref{fig1} for the notation used in this figure. \label{fig3}}
\end{center}
\end{figure}

Considering a model with $x_{\p} = 0.1$ and a purely poloidal 
magnetic field, we show in figure \ref{fig2} the scaling of the polar-led Alfv\'en modes 
 with the entrainment $\veps_\star$. We explore a wide parameter space: 
from the fully-entrained `single-fluid' limit, to entrainment values expected in the neutron star's core, passing 
by the unentrained case  $\veps_{\star}=1$. The agreement with the expected scaling
$\sigma\propto\veps_\star^{1/2}$ is again very good. 
The single-fluid results can be obtained in the strong entrainment
limit with large effective mass $m_{\x}^{\star} \gg m_{\x}$,  which leads to $\veps_{\star} \simeq x_{\p}$ and 
thus to $\sigma \simeq \sigma_0$, see equation~(\ref{eq:pw}). 

The typical effective masses expected in a neutron star core without strong pinning range in the interval 
$ 0.93 \lesssim m_{\n}^{\star} / m_{\n}  \lesssim 1$ and $ 0.4  \lesssim m_{\p}^{\star} / m_{\p}  \lesssim 0.95$~\citep{2008MNRAS.388..737C}.  
These values, which lead to  $ 1 \lesssim \veps_{\star}^{1/2}  \lesssim 1.7$,  may increase the mode frequencies with respect to the single-fluid case by a factor of five in a star 
with $x_{\p}=0.1$.  
It is then natural to wonder whether this strong effect on the spectrum may be influenced by the presence of an elastic crust. 
The conditions in the inner crust, where   we expect 
a lattice of ions permeated by a gas of superfluid neutrons, can be  in fact  very different. Recent calculations show that 
 superfluid neutrons  may be efficiently  entrained by nuclei due to  Bragg scattering, and that the neutron effective mass at the bottom 
of the crust can be  large as $m_{\n}^{\star} \simeq 14 m_{\n}$~\citep{2012PhRvC..85c5801C}. 
This strong entrainment can limit the relative two-fluid motion and produces a 10\% correction 
of the shear mode frequencies determined with a single-fluid model~\citep{2009MNRAS.396..894A,2009CQGra..26o5016S,2012MNRAS.419..638P, 2012arXiv1210.0955S}. 
However, the crust's entrainment  may be less relevant for the Alfv\'en modes due to their 
global nature. A rough estimate can be determined by taking  the effective masses of the crust and the core calculated 
by~\cite{2005NuPhA.748..675C} and \cite{2005NuPhA.747..109C,2006NuPhA.773..263C,2008MNRAS.388..737C}.  
If we insert these values in the unstratified model with $x_{\p}=0.1$ used by~\cite{2012MNRAS.419..638P} and take an average 
of the entrainment profile, we obtain $\sqrt{ \langle \veps_{\star} \rangle } = 1.2$. 
In our purely two-fluid star this value would increase the Alfv\'en mode frequencies  by a factor of 3.8 with respect to the single-fluid case. 
This estimate must be considered with caution, as the coupling of the magnetic field and the crust may lead to a different dynamical evolution and results. 
This issue will be addressed in a future paper.

When the star has a purely toroidal magnetic field we are able to evolve  stably  both 
axial and polar initial data and extract mode frequencies. In figure 
\ref{fig3} we show the effects of proton fraction 
on the axial- and polar-led Alfv\'en mode frequencies for a star 
with a toroidal field and zero entrainment. 
In this case too the scaling of the mode frequencies is well approximated by 
equation~(\ref{eq:pw}) and we are able to recover the results of
single-fluid magnetised models. 

From figures \ref{fig1} and \ref{fig3} we are able to compare the
effect of different magnetic field geometries on the oscillation
spectrum of a magnetar. Let us concentrate on the results for
$x_\p^{-1/2}\approx 3.2$, i.e. $x_\p=0.1$, and just the polar-led
modes (ignoring the single axial-led mode in figure \ref{fig3}). For a
background poloidal field we find five widely-spaced Alfv\'en
modes in the broad range $50-500$ Hz. By contrast, for a purely
toroidal field there are only three modes, in the narrower interval
$180-320$ Hz.  These differences suggest the possibility of
constraining the field geometry with future QPO observations.

Finally, in a star with a toroidal field we find the
Alfv\'en modes again scale in the expected manner with entrainment
($\sigma\propto\veps_\star^{1/2}$), as for the poloidal-field
case. Since the plot of this contains no new information, however, we 
omit it for brevity.

\subsection{Oscillations of stratified magnetars}

We now turn our attention to the oscillation spectrum of stratified models and focus on  purely toroidal magnetic fields. 
The poloidal configurations are currently less numerically stable, and
we want to return to these at a later date. We use a weaker magnetic
field than in the previous subsection here, with an average strength
of $5\times 10^{15}$ G --- again for reasons of numerical stability.

\begin{figure}
\begin{center}
\includegraphics[height=75mm]{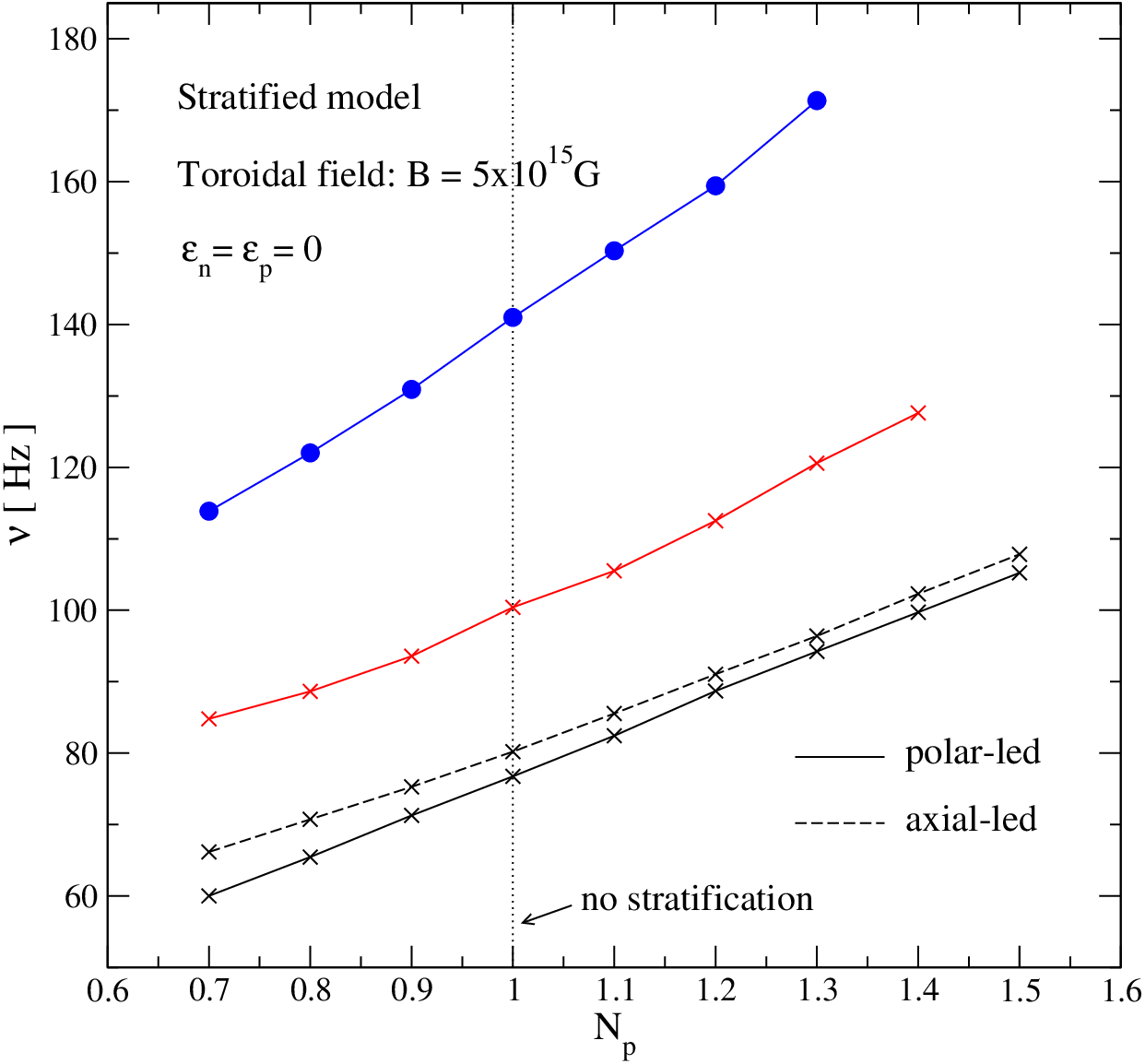}
\caption{ Axial- and polar-led $m=2$ Alfv\'en modes for a series of stratified
  stars with purely toroidal magnetic field with $B=5\times 10^{15}$~G, zero entrainment and $N_{\n} = 1$. 
 The horizontal axis displays the proton polytropic index  $N_{\p}$ which is related to the degree of stratification. 
  Stratification is zero for  $N_{\p} = 1$ and increases for values of $N_{\p}$ much different from $N_{\n}$. 
 The central proton fraction is $x_{\p}=0.15$  for all these stellar models.  On the vertical axis we show 
  the mode frequency in physical units for a star with $M=1.4 M_{\odot}$ and $R=10\, \textrm{km}$. 
  The curves are represented with the same notation used in figure~\ref{fig1}.
  \label{fig4}}
\end{center}
\end{figure}

The effects of stratification can be studied by using equation~(\ref{eq:EosPR}) with 
$N_{\n} \neq N_{\p}$. We consider a sequence of stars with various composition gradients by 
setting  $N_{\n}=1$ and exploring a range of proton polytropic indices, $0.7\leq N_\p\leq 1.5$. 
For this range we are able to run the code for a sufficient evolution time to extract  mode frequencies reliably 
(15+ Alfv\'en times). The central proton fraction is $x_{\p}=0.15$ for all these models and tends to 
zero (unity) for $N_{\p} >1$ ($N_{\p} < 1$). The frequencies of the axial- and polar-led $m=2$ Alfv\'en modes are 
shown in figure~\ref{fig4} for stars with toroidal field $B=5\times10^{15}$~G and zero entrainment. 
The mode frequencies $\nu = \sigma / ( 2 \pi )$ are given in physical units for a star with $M=1.4 M_{\odot}$ 
and  $R=10\, \textrm{km}$. Note that in this stratified stellar sequence the central mass density $\rho_0$ does not scale linearly with $N_{\p}$. 
Therefore, the dimensionless mode frequencies cannot be  shown in figure~\ref{fig4}  together with the physical values.  

Our results show that composition gradients may significantly affect the Alfv\'en modes with respect to unstratified models. 
The $m=2$ modes at $N_{\p} = 1.5$ are a factor of about 1.38 larger than the unstratified case $N_{\p}=1$.  As shown 
recently by  \citet{2012MNRAS.419..732L}, a ``realistic'' proton gradient can be reproduced 
by the EoS~(\ref{eq:EosPR})  with a proton-fluid index close to $N_{\p}=2$.  Unfortunately,  
this value is slightly beyond our current numerical capacity. However, considering a linear fit of the 
mode frequencies shown in figure~\ref{fig4} we find that a model with
$N_{\p}=2$ should oscillate with frequencies roughly 1.67 larger than the $N_{\p}=1$ case. For instance, the linear fit 
for the polar-led Alfv\'en mode which  has the lowest frequencies in figure~\ref{fig4} is given by 
\begin{equation} 
\nu = (20.0 + 56.9N_{\p}) \textrm{ Hz.} \label{eq:fit}
\end{equation}

From the unstratified model we can identify the modes already calculated by~\citet{2010MNRAS.405..318L}    in the single-fluid limit. 
In figure~\ref{fig4} they are represented with a cross.

\section{Discussion} \label{sec:discussion}

The main result of this work is that the multi-fluid physics of a
magnetar is likely to result in considerably higher Alfv\'en-mode
frequencies than a standard single-fluid barotropic model would
indicate. Although this was predicted at a qualitative level by an
earlier plane-wave analysis~\citep{2009MNRAS.396..894A}, this paper is the first quantitative study of
the oscillation spectrum  of a multi-fluid magnetar.   

We have studied the time evolution of non-axisymmetric oscillations of 
purely two-fluid stars with superfluid neutrons and normal (not superconducting) protons. 
The effects of an elastic crust  will be included in future work.  We have considered 
various magnetic field geometries, 
 proton fractions, entrainment  and  composition stratifications and analysed their impact on 
the Alfv\'en mode spectrum.

Starting with unstratified models, we find that realistic values
for the proton fraction and entrainment lead to Alfv\'en mode frequencies 
three to four times larger than the single-fluid results. 
If we move from an unstratified star to one with a more realistic composition gradient the mode frequencies may be increased 
by another factor of $\sim 1.67$. 
From these results we can extrapolate that a `typical' multi-fluid magnetar 
could have QPO frequencies which are roughly a factor of five/six higher than
expected from a single-fluid unstratified model. Based on our results  and using equations~(\ref{eq:pw}) and~(\ref{eq:fit}), 
we may summarise the
expected scaling of Alfv\'en modes in the presence of superfluid
physics as:
\begin{equation} \label{multifluid_scaling}
\sigma \approx 6.3   \sigma_0 \left[ 0.15+0.85  \left( \frac{N_\p}{2} \right) \right]  \brac{\frac{\veps_\star}{1.3}}^{1/2} \brac{\frac{x_\p(0)}{0.1}}^{-1/2}   ,
\end{equation}
where $\sigma$ is a mode of a multi-fluid magnetar,  $\sigma_0$ is the
corresponding single-fluid mode, and  $x_\p(0)$ is the central proton
fraction. 

Furthermore, our results suggest that there is some hope for constraining the
magnetic field geometry of a magnetar from the QPO frequency distribution.
For a typical unstratified stellar model with a
poloidal field we find five Alfv\'en modes, in the wide range $50-500$
Hz; instead, for a toroidal field there are three modes in a far
narrower range, $180-320$ Hz.

Let us consider a magnetar with average field strength
$B=10^{16}$G. Although this seems rather high, it
corresponds to a more reasonable polar-cap value of $5\times 10^{15}$ G if
we assume a poloidal-field geometry; if the field is more `buried' (e.g. with a strong
interior toroidal field) then it may have an average strength of
$10^{16}$ G but a polar-cap value of only $10^{15}$ G.  For this field
strength, typical single-fluid
lowest-order Alfv\'en mode frequencies are roughly\footnote{This
  refers to $m=0$ axial modes; \citet{2009MNRAS.395.1163S} finds a
  fundamental polar $m=0$ mode at $300$ Hz.} $10-50$ Hz for
$m=0$ and $50-150$ Hz for $m=2$; higher $l$-multipoles or values of $m$
produce higher frequencies
\citep{2008MNRAS.385L...5S,2009MNRAS.396.1441C,2010MNRAS.405..318L,2011MNRAS.412.1730L,2012MNRAS.421.2054G}.
Temporarily ignoring the various simplifications in our model, let us
look at the repercussions of formula \eqref{multifluid_scaling} for
identification of magnetar QPOs. The strong
150 Hz QPO of SGR 1806-20 or the 155 Hz QPO of SGR 1900+14 may be
interpreted as an axisymmetric Alfv\'en QPO of the magnetar's core ---
the multi-fluid equivalent of a 25-Hz peak predicted by single-fluid
models. Similarly, the long-lived 625 Hz peak of SGR 1806-20 would
then be the multi-fluid equivalent of a 100-Hz single-fluid mode, which is in the
range expected for an $m=2$ Alfv\'en mode.

The above discussion does not provide an interpretation for the lower-frequency
magnetar QPOs and neglects the effect of the crust.
The effect of superfluidity on the crustal shear modes is similar to what is described in this paper for the Alfv\'en modes. However, 
a realistic entrainment~\citep{2012PhRvC..85c5801C} may have a very different
quantitative impact on these two classes of modes.  
In fact, the effective mass of neutrons may be quite large at the
bottom of the crust leading to a total correction of the shear mode of about 10\% with respect 
to the single-fluid models. The Alfv\'en modes however may be less affected by this entrainment configuration which is confined only in a limited region of the star. 
In conclusion, multi-fluid dynamics with a realistic entrainment may
increase significantly the Alfv\'en modes as described in the last
paragraph, but only produce a 10 percent correction to the 
shear modes. 
Therefore, we suggest that the observed magnetar QPOs 
may fall into two classes: below roughly $50$ Hz they may be magneto-elastic
crustal modes, whilst above this value they could represent Alfv\'en oscillations
of the multi-fluid core.

The attractive feature of our interpretation is that it does not rely
on the long-term excitation of high multipoles  of the magnetar's
oscillation modes, but instead suggests that observed QPOs are all
low-order modes originating from the core or crust. Nonetheless, our
description of a magnetar is still rather approximate. 
We must refine
our superfluid magnetised models and also include an elastic crust.
One key issue is whether magnetically-modified elastic
modes are really able to explain all the lower-frequency
QPOs.  In the axisymmetric case, two-fluid physics may move the Alfv\'en continuum to higher
frequencies. In this way, more crustal modes could be outside the
continuum and live for a longer time. Alternatively, some
low-frequency QPOs could still be core modes if the
magnetic field is weaker than expected, or if the two-fluid
enhancement of axisymmetric oscillations is less significant than
predicted by equation \eqref{multifluid_scaling}. 

Our conclusions are based on a multi-fluid star where the dynamics of neutrons and
protons is essentially decoupled at linear perturbation order, interacting only through the entrainment.
We assume that the magnetar's internal field
strength is above the critical value at which superconductivity is
destroyed and hence that the protons form a normal fluid, but this is
far from certain.  If instead the protons form a type-II
superconductor, the oscillation spectrum will certainly be
affected. The nature of the oscillations changes and their frequency
is altered by a factor of $\sqrt{H_{c1}/B}$, where $H_{c1}\approx
10^{15}$ G \citep{1998MNRAS.296..903M}. In addition, the presence of a
magnetic force on the neutrons and coupling between proton fluxtubes and
neutron vortices could lower the frequency of oscillations with
respect to our results
\citep{2008MNRAS.391..283V,2011MNRAS.410..805G}. These important
issues deserve more quantitative attention in future studies of magnetar QPOs.

\section*{Acknowledgements}

We acknowledge support from the German Science Foundation (DFG) via SFB/TR7. 
We are also pleased to thank A. Colaiuda,  K. Glampedakis and K. Kokkotas for fruitful discussions.

\appendix

\nocite*

\begin{thebibliography}{}

\bibitem[\protect\citeauthoryear{{Andersson}, {Comer} \& {Grosart}}{{Andersson}
  et~al.}{2004}]{2004MNRAS.355..918A}
{Andersson} N.,  {Comer} G.~L.,    {Grosart} K.,  2004, \mnras, 355, 918

\bibitem[\protect\citeauthoryear{{Andersson}, {Comer} \&
  {Langlois}}{{Andersson} et~al.}{2002}]{2002PhRvD..66j4002A}
{Andersson} N.,  {Comer} G.~L.,    {Langlois} D.,  2002, \prd, 66, 104002

\bibitem[\protect\citeauthoryear{{Andersson}, {Glampedakis} \&
  {Samuelsson}}{{Andersson} et~al.}{2009}]{2009MNRAS.396..894A}
{Andersson} N.,  {Glampedakis} K.,    {Samuelsson} L.,  2009, \mnras, 396, 894

\bibitem[\protect\citeauthoryear{{Barat}, {Hayles}, {Hurley}, {Niel},
  {Vedrenne}, {Desai}, {Kurt}, {Zenchenko} \& {Estulin}}{{Barat}
  et~al.}{1983}]{1983A&A...126..400B}
{Barat} C.,  {Hayles} R.~I.,  {Hurley} K.,  {Niel} M.,  {Vedrenne} G.,  {Desai}
  U.,  {Kurt} V.~G.,  {Zenchenko} V.~M.,    {Estulin} I.~V.,  1983, \aap, 126,
  400

\bibitem[\protect\citeauthoryear{{Baym}, {Pethick} \& {Pines}}{{Baym}
  et~al.}{1969}]{1969Natur.224..673B}
{Baym} G.,  {Pethick} C.,    {Pines} D.,  1969, \nat, 224, 673

\bibitem[\protect\citeauthoryear{{Carter}, {Chamel} \& {Haensel}}{{Carter}
  et~al.}{2005}]{2005NuPhA.748..675C}
{Carter} B.,  {Chamel} N.,    {Haensel} P.,  2005, Nuclear Physics A, 748, 675

\bibitem[\protect\citeauthoryear{{Chamel}}{{Chamel}}{2005}]{2005NuPhA.747..109C}
{Chamel} N.,  2005, Nuclear Physics A, 747, 109

\bibitem[\protect\citeauthoryear{{Chamel}}{{Chamel}}{2006}]{2006NuPhA.773..263C}
{Chamel} N.,  2006, Nuclear Physics A, 773, 263

\bibitem[\protect\citeauthoryear{{Chamel}}{{Chamel}}{2008}]{2008MNRAS.388..737C}
{Chamel} N.,  2008, \mnras, 388, 737

\bibitem[\protect\citeauthoryear{{Chamel}}{{Chamel}}{2012}]{2012PhRvC..85c5801C}
{Chamel} N.,  2012, \prc, 85, 035801

\bibitem[\protect\citeauthoryear{{Colaiuda}, {Beyer} \& {Kokkotas}}{{Colaiuda}
  et~al.}{2009}]{2009MNRAS.396.1441C}
{Colaiuda} A.,  {Beyer} H.,    {Kokkotas} K.~D.,  2009, \mnras, 396, 1441

\bibitem[\protect\citeauthoryear{{Colaiuda} \& {Kokkotas}}{{Colaiuda} \&
  {Kokkotas}}{2012}]{2012MNRAS.423..811C}
{Colaiuda} A.,  {Kokkotas} K.~D.,  2012, \mnras, 423, 811

\bibitem[\protect\citeauthoryear{{Duncan}}{{Duncan}}{1998}]{1998ApJ...498L..45D}
{Duncan} R.~C.,  1998, \apjl, 498, L45

\bibitem[\protect\citeauthoryear{{Gabler}, {Cerd{\'a}-Dur{\'a}n},
  {Stergioulas}, {Font} \& {M{\"u}ller}}{{Gabler}
  et~al.}{2012}]{2012MNRAS.421.2054G}
{Gabler} M.,  {Cerd{\'a}-Dur{\'a}n} P.,  {Stergioulas} N.,  {Font} J.~A.,
  {M{\"u}ller} E.,  2012, \mnras, 421, 2054

\bibitem[\protect\citeauthoryear{{Glampedakis}, {Andersson} \&
  {Samuelsson}}{{Glampedakis} et~al.}{2011}]{2011MNRAS.410..805G}
{Glampedakis} K.,  {Andersson} N.,    {Samuelsson} L.,  2011, \mnras, 410, 805

\bibitem[\protect\citeauthoryear{{Glampedakis}, {Samuelsson} \&
  {Andersson}}{{Glampedakis} et~al.}{2006}]{2006MNRAS.371L..74G}
{Glampedakis} K.,  {Samuelsson} L.,    {Andersson} N.,  2006, \mnras, 371, L74

\bibitem[\protect\citeauthoryear{{Ho}, {Glampedakis} \& {Andersson}}{{Ho}
  et~al.}{2012}]{2012MNRAS.422.2632H}
{Ho} W.~C.~G.,  {Glampedakis} K.,    {Andersson} N.,  2012, \mnras, 422, 2632

\bibitem[\protect\citeauthoryear{{Israel}, {Belloni}, {Stella}, {Rephaeli},
  {Gruber}, {Casella}, {Dall'Osso}, {Rea}, {Persic} \& {Rothschild}}{{Israel}
  et~al.}{2005}]{2005ApJ...628L..53I}
{Israel} G.~L.,  {Belloni} T.,  {Stella} L.,  {Rephaeli} Y.,  {Gruber} D.~E.,
  {Casella} P.,  {Dall'Osso} S.,  {Rea} N.,  {Persic} M.,    {Rothschild}
  R.~E.,  2005, \apjl, 628, L53

\bibitem[\protect\citeauthoryear{{Jones}, {Andersson} \& {Stergioulas}}{{Jones}
  et~al.}{2002}]{2002MNRAS.334..933J}
{Jones} D.~I.,  {Andersson} N.,    {Stergioulas} N.,  2002, \mnras, 334, 933

\bibitem[\protect\citeauthoryear{{Lander}, {Andersson} \&
  {Glampedakis}}{{Lander} et~al.}{2012}]{2012MNRAS.419..732L}
{Lander} S.~K.,  {Andersson} N.,    {Glampedakis} K.,  2012, \mnras, 419, 732

\bibitem[\protect\citeauthoryear{{Lander} \& {Jones}}{{Lander} \&
  {Jones}}{2011}]{2011MNRAS.412.1730L}
{Lander} S.~K.,  {Jones} D.~I.,  2011, \mnras, 412, 1730

\bibitem[\protect\citeauthoryear{{Lander} \& {Jones}}{{Lander} \&
  {Jones}}{2012}]{2012MNRAS.tmp.3162L}
{Lander} S.~K.,  {Jones} D.~I.,  2012, \mnras, p.~3162

\bibitem[\protect\citeauthoryear{{Lander}, {Jones} \& {Passamonti}}{{Lander}
  et~al.}{2010}]{2010MNRAS.405..318L}
{Lander} S.~K.,  {Jones} D.~I.,    {Passamonti} A.,  2010, \mnras, 405, 318

\bibitem[\protect\citeauthoryear{{Levin}}{{Levin}}{2006}]{2006MNRAS.368L..35L}
{Levin} Y.,  2006, \mnras, 368, L35

\bibitem[\protect\citeauthoryear{{Levin}}{{Levin}}{2007}]{2007MNRAS.377..159L}
{Levin} Y.,  2007, \mnras, 377, 159

\bibitem[\protect\citeauthoryear{{Lockitch} \& {Friedman}}{{Lockitch} \&
  {Friedman}}{1999}]{1999ApJ...521..764L}
{Lockitch} K.~H.,  {Friedman} J.~L.,  1999, \apj, 521, 764

\bibitem[\protect\citeauthoryear{{Markey} \& {Tayler}}{{Markey} \&
  {Tayler}}{1973}]{1973MNRAS.163...77M}
{Markey} P.,  {Tayler} R.~J.,  1973, \mnras, 163, 77

\bibitem[\protect\citeauthoryear{{Mendell}}{{Mendell}}{1998}]{1998MNRAS.296..903M}
{Mendell} G.,  1998, \mnras, 296, 903

\bibitem[\protect\citeauthoryear{{Passamonti} \& {Andersson}}{{Passamonti} \&
  {Andersson}}{2011}]{2011MNRAS.413...47P}
{Passamonti} A.,  {Andersson} N.,  2011, \mnras, 413, 47

\bibitem[\protect\citeauthoryear{{Passamonti} \& {Andersson}}{{Passamonti} \&
  {Andersson}}{2012}]{2012MNRAS.419..638P}
{Passamonti} A.,  {Andersson} N.,  2012, \mnras, 419, 638

\bibitem[\protect\citeauthoryear{{Passamonti}, {Haskell} \&
  {Andersson}}{{Passamonti} et~al.}{2009}]{2009MNRAS.396..951P}
{Passamonti} A.,  {Haskell} B.,    {Andersson} N.,  2009, \mnras, 396, 951

\bibitem[\protect\citeauthoryear{{Prix} \& {Rieutord}}{{Prix} \&
  {Rieutord}}{2002}]{2002A&A...393..949P}
{Prix} R.,  {Rieutord} M.,  2002, \aap, 393, 949

\bibitem[\protect\citeauthoryear{{Reisenegger}}{{Reisenegger}}{2009}]{2009AA...499..557R}
{Reisenegger} A.,  2009, \aap, 499, 557

\bibitem[\protect\citeauthoryear{{Samuelsson} \& {Andersson}}{{Samuelsson} \&
  {Andersson}}{2009}]{2009CQGra..26o5016S}
{Samuelsson} L.,  {Andersson} N.,  2009, Classical and Quantum Gravity, 26,
  155016

\bibitem[\protect\citeauthoryear{{Sotani} \& {Kokkotas}}{{Sotani} \&
  {Kokkotas}}{2009}]{2009MNRAS.395.1163S}
{Sotani} H.,  {Kokkotas} K.~D.,  2009, \mnras, 395, 1163

\bibitem[\protect\citeauthoryear{{Sotani}, {Kokkotas} \&
  {Stergioulas}}{{Sotani} et~al.}{2008}]{2008MNRAS.385L...5S}
{Sotani} H.,  {Kokkotas} K.~D.,    {Stergioulas} N.,  2008, \mnras, 385, L5

\bibitem[\protect\citeauthoryear{{Sotani}, {Nakazato}, {Iida} \&
  {Oyamatsu}}{{Sotani} et~al.}{2012}]{2012arXiv1210.0955S}
{Sotani} H.,  {Nakazato} K.,  {Iida} K.,    {Oyamatsu} K.,  2012,
  arXiv:1210.0955

\bibitem[\protect\citeauthoryear{{Strohmayer} \& {Watts}}{{Strohmayer} \&
  {Watts}}{2005}]{2005ApJ...632L.111S}
{Strohmayer} T.~E.,  {Watts} A.~L.,  2005, \apjl, 632, L111

\bibitem[\protect\citeauthoryear{{van Hoven} \& {Levin}}{{van Hoven} \&
  {Levin}}{2008}]{2008MNRAS.391..283V}
{van Hoven} M.,  {Levin} Y.,  2008, \mnras, 391, 283

\bibitem[\protect\citeauthoryear{{Watts} \& {Strohmayer}}{{Watts} \&
  {Strohmayer}}{2006}]{2006ApJ...637L.117W}
{Watts} A.~L.,  {Strohmayer} T.~E.,  2006, \apjl, 637, L117

\bibitem[\protect\citeauthoryear{{Wright}}{{Wright}}{1973}]{1973MNRAS.162..339W}
{Wright} G.~A.~E.,  1973, \mnras, 162, 339

\end{thebibliography}

\label{lastpage}

\end{document}